\documentclass[fleqn,usenatbib]{mnras}
\usepackage{newtxtext}
\usepackage[T1]{fontenc}
\usepackage{ae,aecompl}
\usepackage{graphicx}    
\usepackage{amsmath}     
\usepackage{amssymb}     

\usepackage{newtxmath}
\usepackage{multicol}    
\usepackage{bm}          
\usepackage{pdflscape}   
\usepackage{xspace}
\newcommand{\Msun}{\,$M_{\sun}$\xspace}
\newcommand{\Rsun}{\,$R_{\sun}$\xspace}
\newcommand{\kms}{\,km\,s$^{-1}$\xspace}
\newcommand{\ergs}{\,erg\,s$^{-1}$\xspace}
\newcommand{\Ha}{H$\alpha$\xspace}
\newcommand{\Hb}{H$\beta$\xspace}
\newcommand{\HeI}{\ion{He}{i}\xspace}
\newcommand{\OI}{\ion{O}{i}\xspace}
\newcommand{\NaI}{\ion{Na}{i}\xspace}
\newcommand{\MgI}{\ion{Mg}{i}\xspace}
\newcommand{\FeII}{\ion{Fe}{ii}\xspace}
\newcommand{\A}{\,\AA\xspace}
\title[Enormous explosion energy of SN~2017gmr] 
   {Enormous explosion energy of Type IIP SN~2017gmr with bipolar $^{56}$Ni
   ejecta}
\author[V. P. Utrobin et al.]{%
Victor P. Utrobin$^{1,2,3}$,\thanks{E-mail: utrobin@itep.ru}
Nikolai N. Chugai$^{2}$,
Jennifer E. Andrews$^{4}$,
Nathan Smith$^{4}$,
\newauthor
Jacob Jencson$^{4}$,
D. Andrew Howell$^{5,6}$,
Jamison Burke$^{5,6}$,
Daichi Hiramatsu$^{5,6}$,
\newauthor
Curtis McCully$^{5,6}$, and
K. Azalee Bostroem$^{7}$
\\
$^{1}$NRC `Kurchatov Institute' --
      Institute for Theoretical and Experimental Physics,
      B.~Cheremushkinskaya St. 25, 117218 Moscow, Russia \\
$^{2}$Institute of Astronomy, Russian Academy of Sciences, Pyatnitskaya
      St. 48, 119017 Moscow, Russia\\
$^{3}$Max-Planck-Institut f\"ur Astrophysik, Karl-Schwarzschild-Str. 1,
      85748 Garching, Germany \\
$^{4}$Steward Observatory, University of Arizona, 933 North Cherry Avenue,
      Tucson, AZ 85721-0065, USA\\
$^{5}$Department of Physics, University of California, Santa Barbara,
      CA 93106-9530, USA\\
$^{6}$Las Cumbres Observatory, 6740 Cortona Dr, Suite 102, Goleta,
      CA 93117-5575, USA\\
$^{7}$Department of Physics and Astronomy, University of California,
      1 Shields Avenue, Davis, CA 95616-5270, USA\\
}

\date{Accepted 2021 May 9. Received 2021 April 19; in original form 2021 March 1}

\pubyear{2021}

\begin{document}
\label{firstpage}
\pagerange{\pageref{firstpage}--\pageref{lastpage}}
\maketitle
%
\begin{abstract}
The unusual Type IIP SN~2017gmr is revisited in order to pinpoint
   the origin of its anomalous features, including the peculiar light curve
   after about 100\,days.
The hydrodynamic modelling suggests the enormous explosion energy of
   $\approx$10$^{52}$\,erg.
We find that the light curve with the prolonged plateau/tail transition can
   be reproduced either in the model with a high hydrogen abundance in
   the inner ejecta and a large amount of radioactive $^{56}$Ni, or in
   the model with an additional central energy source associated with
   the fallback/magnetar interaction in the propeller regime.
The asymmetry of the late \Ha emission and the reported linear polarization
   are reproduced by the model of the bipolar $^{56}$Ni ejecta.
The similar bipolar structure of the oxygen distribution is responsible for
   the two-horn structure of the [\OI] 6360, 6364\A emission.
The bipolar $^{56}$Ni structure along with the high explosion energy are
   indicative of the magneto-rotational explosion. 
We identify narrow high-velocity absorption features in \Ha and \HeI 10830\A
   lines with their origin in the fragmented cold dense shell formed due to
   the outer ejecta deceleration in a confined circumstellar shell.
\end{abstract}
\begin{keywords}
hydrodynamics -- methods: numerical -- supernovae: general --
supernovae: individual: SN~2017gmr
\end{keywords}

\section{Introduction} 
\label{sec:intro}
Type IIP supernovae (SNe~IIP) originate from a core collapse of massive stars
   ($>$9\Msun) that retain a significant fraction of the hydrogen envelope
   until the explosion.
The general paradigm is that the SN~IIP light curve at the plateau stage is
   maintained by the release of the internal energy deposited during the
   propagation of the shock wave through the presupernova (pre-SN) envelope
   \citep{GIN_71}, whereas the luminosity tail is powered by
   the radioactive decay $^{56}$Co $\to$ $^{56}$Fe \citep{WW_80}.
The debatable explosion mechanisms of SNe~IIP or, in a broad sense, of
   core-collapse supernovae include the neutrino-driven explosion 
   \citep{CW_66, Jan_17, BV_21}, the magneto-rotational explosion related
   to the magnetar formation \citep{LW_70, BK_71, KHO_99}, and the jet-powered
   supernovae related to the collapsar (rotating black hole plus disk)
   formation \citep{MWH_01}.

In most observed cases one cannot distinguish between different options,
   because the outcome of the SN explosion with the typical energy of
   $\sim$10$^{51}$\,erg is not sensitive to the explosion mechanism.
An exception is the case when the explosion energy inferred from the 
   hydrodynamic modelling significantly exceeds the upper limit for the
   neutrino-driven mechanism $E_{up} \approx 2\times10^{51}$\,erg
   \citep{Jan_17}.
Among eleven SNe~IIP explored so far via the uniform hydrodynamic modelling
   \citep[cf.][]{UC_19} only two events can be attributed to the category of
   high-energy SNe~IIP with $E > E_{up}$: SN~2009kf with
   $E = 2.15\times10^{52}$\,erg \citep{UCB_10} and SN~2000cb with
   $E = 4.4\times10^{51}$\,erg \citep{UC_11}.  

The recent Type IIP SN~2017gmr is another candidate for
   high-energy SNe~IIP, because it exhibits both high luminosity and
   high expansion velocities \citep{ASV_19}.
The hydrodynamic modelling of SN~2017gmr \citep{GB_20} prefers a high
   explosion energy of $\approx$5$\times10^{51}$\,erg despite of the claimed
   model degeneracy with respect to SN parameters.
It is noteworthy that SN~2017gmr shows signatures of asymmetry in the \Ha and
   [\OI] 6360, 6364\A emission lines on day 312 \citep{ASV_19} 
   and in the polarization \citep{NCP_19}, which indicate a non-spherical 
   explosion.

Even more remarkable feature of SN~2017gmr is a steeper luminosity decline of
   the post-plateau tail compared to the $^{56}$Co decay luminosity
   \citep{ASV_19}.
The hydrodynamic modelling of \citet{GB_20} is unable to account for this
   behavior that is interpreted as an excess in the observed luminosity over
   the computed luminosity at the radioactive tail.
Possible explanations for the fast tail decline include the early escape of
   gamma-quanta due to the decay of the radioactive $^{56}$Co,
   the circumstellar (CS) interaction, or the early dust formation
   \citep{ASV_19}, although the latter seems unlikely because of
   the high gas temperature at this stage.
The CS interaction might be relevant, since early spectra show the narrow
   \Ha emission with broad wings indicating the dense CS shell.
Another possibility might be the high velocity of radioactive $^{56}$Ni,
   favouring the early escape of the gamma rays similar to the case of
   SN~2013ej \citep{UC_17}.
The applicability of this conjecture for SN~2017gmr at the moment is unclear
   and must be examined.

Here we address unusual features of SN~2017gmr with an emphasis on the fast
   decline of the luminosity tail.
To this end we revisit the construction of the bolometric light curve
   (Section~\ref{sec:bolcrv}) and then perform the hydrodynamic modelling
   and examine a number of possibilities for the origin of the fast tail
   decline (Section~\ref{sec:hydro}).
The asymmetry effects in emission lines at the nebular stage and
   the polarization are explored in order to constrain the extent of
   mixing of radioactive $^{56}$Ni (Section~\ref{sec:asym}). 
We then address the effects of the CS interaction at the photospheric
   stage (Section~\ref{sec:csi}).
Finally, results are summarized and discussed in Section~\ref{sec:disc}.

Below we adopt the distance of 19.6\,Mpc and the reddening of
   $A(B-V) = 0.3$\,mag \citep{ASV_19}.
The explosion date is set to be 2017 September 1.88, MJD 57997.89, which is
   recovered from the fit of the earliest $r$ magnitudes by the hydrodynamic
   modelling.
This moment is 1.20 day earlier compared to that adopted by \citet{ASV_19}.

\section{Bolometric light curve and photospheric velocities}
\label{sec:bolcrv}
%
\begin{figure}
   \includegraphics[width=\columnwidth, clip, trim=0 20 47 176]{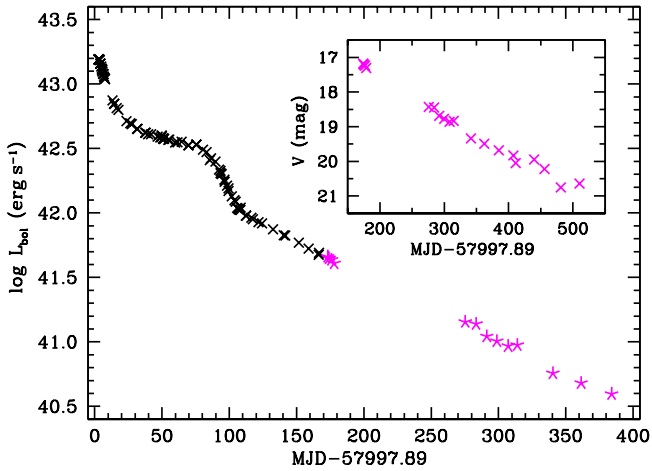}
   \caption{
   Bolometric light curve integrated from NUV to NIR with blackbody corrections 
      (\emph{gray crosses\/}) and the late time reconstruction based on $V$
      magnitudes (\emph{magenta asterisks\/}).
   Inset shows late time $V$ magnitudes.
   }
   \label{fig:blcrv}
\end{figure}
\begin{figure}
   \includegraphics[width=\columnwidth, clip, trim=8 17 46 250]{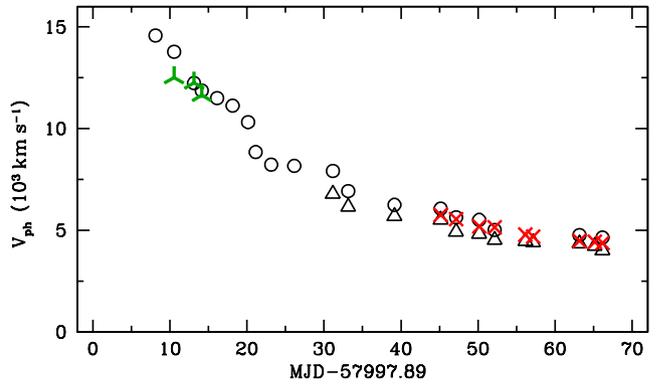}
   \caption{
   Velocity at the photosphere level of SN~2017gmr inferred from absorption
      minima of four lines: \Hb (\emph{black circles\/}), \HeI 5876\A
      (\emph{green three point star symbols\/}), \FeII 5169\A
      (\emph{black triangles\/}), and \NaI 5892\A doublet
      (\emph{red crosses\/}).
   }
   \label{fig:vphcrv}
\end{figure}

Additional late-time photometry presented here and not published in
   \citet{ASV_19} was obtained via the Las Cumbres Observatory 1-m telescope
   network \citep{BBB_13} with the Sinistro cameras in the framework of
   the Global Supernova Project.
The data were reduced using the Beautiful Algorithms to Normalize Zillions of
   Astronomical Images (BANZAI) pipeline \citep{MVH_18}.
PSF-fitting photometry was then performed using {\sc lcogtsnpipe}
   \citep{VHS_16}, a PyRAF-based photometric reduction pipeline.
$UBV$-band data were calibrated to Vega magnitudes \citep{Ste_00} using
   standard fields observed on the same night with the same telescope,
   while $gri$-band data were calibrated to AB magnitudes using
   the Sloan Digital Sky Survey \citep[SDSS Collaboration;][]{SDSS_17}.

As described in \citet{ASV_19}, a quasi-bolometric light curve was created
   using the routine \textsc{superbol} \citep{Nic_18}, where the reddening
   and redshift corrected photometry in each band (from $UV$ to $IR$) was
   interpolated with the $g$-band as reference, then converted to a spectral
   luminosity (L$_{\lambda}$).
The bolometric luminosity was then computed from the integration of the SED
   for each epoch.
While uncertainties in the reddening ($E(B-V)$ = 0.30 $\pm$ 0.06\,mag) and
   distance (19.6 $\pm$ 1.4\,Mpc) introduce uncertainties in the total
   bolometric luminosity, the bolometric light curve falls much more rapidly
   than expected for a fully-trapped $^{56}$Co decay.

To get an idea of the late-time ($t > 170$\,days) bolometric luminosity, we 
   convert $V$ magnitudes into bolometric fluxes using a relation between
   $V$ and bolometric magnitudes for SN~1987A \citep{CWF_88, WCM_88}.
This procedure suggests that spectral energy distributions of SN~2017gmr and
   SN~1987A are similar and differ only by a constant factor that can be
   fixed by matching bolometric luminosities of SN~2017gmr around day 170
   (Fig.~\ref{fig:blcrv}).

The velocities at the photosphere (Fig.~\ref{fig:vphcrv}) are estimated
   from absorption minima of \Hb, \FeII 5169\A, \HeI 5876\A, and
   \NaI 5892\A doublet in spectra of SN~2017gmr \citep{ASV_19}.
In the case of \NaI we treat photon scatterings in the doublet by the
   Monte Carlo technique to recover the photospheric velocity from
   the absorption minimum.
According to the \FeII 5169\A absorption the photospheric velocity is
   lower by $\sim$500\kms compared to the velocities from \Hb and
   \NaI 5892\A doublet (Fig.~\ref{fig:vphcrv}).
The possible reason for that might be the presence of \MgI 5167, 5173,
   5184\A triplet.
The optical depth of the strongest triplet line \MgI 5184\A is comparable
   to that of \FeII 5169\A for the ionization fraction \MgI/Mg of
   $\sim$0.025.
We performed Monte Carlo simulations of the radiation transfer for
   the blend of \FeII and \MgI triplet assuming that the Sobolev optical
   depth of the strongest \MgI line 5184\A line is equal to that of
   the \FeII 5169\A line.
The absorption minimum in this case turns out shifted towards red by about
   500\kms, which could explain the lower velocity according to the
   \FeII 5169\A line.

Noteworthy, the radial velocity of an absorption minimum ($|v_\mathrm{min}|$)
   of the P Cygni profile may differ from the photospheric velocity
   ($v_\mathrm{ph}$).
The equality $|v_\mathrm{min}| = v_\mathrm{ph}$ takes place only for
   the scattering lines of a moderate strength.
For a strong line even with the conservative scattering the minimum is displaced
   towards blue due to the scattered emission, so in this case
   $|v_\mathrm{min}| > v_\mathrm{ph}$.
This effect is strengthened by a net emission likewise in the case of \Ha.
For a weak absorption, on the contrary, $|v_\mathrm{min}| < v_\mathrm{ph}$,
   because in this case the absorption is formed by a narrow layer close to
   the photosphere and projection effects shift the absorption minimum towards
   zero radial velocity.
We therefore admit that the velocity recovered from the shallow \Hb and \HeI
   absorptions at the early stage ($t < 20$\,days) likely underestimate
   the photospheric velocity by about 20\%.
At the later stage ($t > 20$\,days) the \Hb absorption is moderately strong
   and the net and scattered emissions are suppressed because of the \Hb quanta
   conversion into P$\alpha$ and \Ha.
We thus believe that the \Hb absorption provides a reliable measure of the
   photospheric velocity at this stage.

\section{Hydrodynamic modelling}
\label{sec:hydro}
\subsection{Model overview}
\label{sec:hydro:model}
%
\begin{figure}
   \includegraphics[width=\columnwidth, clip, trim=0 237 54 138]{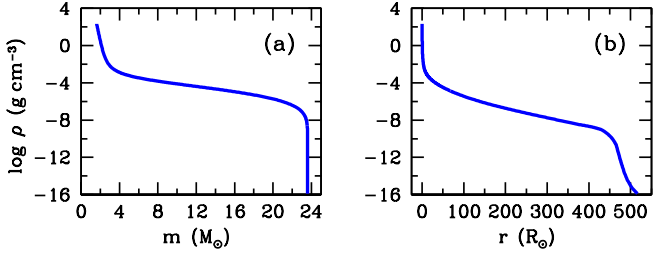}
   \caption{
   Density distribution as a function of interior mass (Panel (a)) and
      radius (Panel (b)) for the pre-SN model HM-optm (Table~\ref{tab:hydmod}).
   The central core of 1.6\Msun is omitted.
   }
   \label{fig:denmr}
\end{figure}
\begin{figure}
   \includegraphics[width=\columnwidth, clip, trim=8 17 46 250]{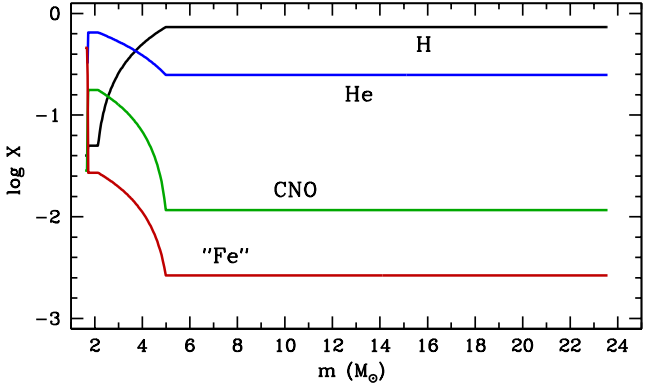}
   \caption{
   Mass fractions of major constituents: hydrogen (\emph{black line\/}),
      helium (\emph{blue line\/}), CNO elements (\emph{green line\/}),
      and Fe-peak elements without radioactive $^{56}$Ni (\emph{red line\/})
      in the ejected envelope of model HM-optm (Table~\ref{tab:hydmod}).
   }
   \label{fig:chcom}
\end{figure}
\begin{figure}
   \includegraphics[width=\columnwidth, clip, trim=15 18 47 208]{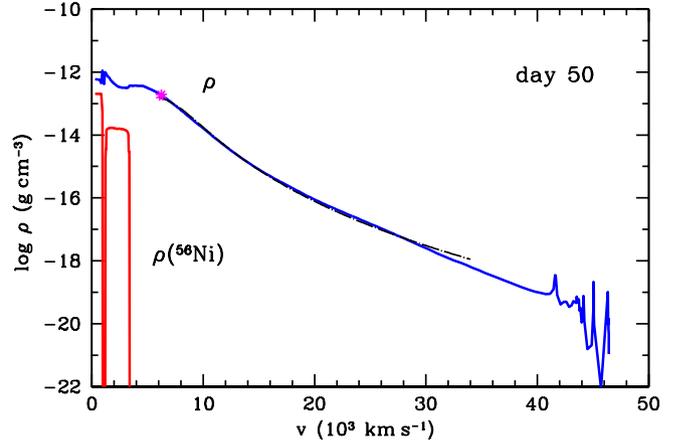}
   \caption{
   The gas (\emph{blue line\/}) and $^{56}$Ni (\emph{red line\/}) density
      as a function of velocity on day 50 for model HM-optm
      (Table~\ref{tab:hydmod}).
   Magenta star indicates the position of the photosphere.
   \emph{Dash-dotted line\/} is the power law $\rho \propto v^{-8}$.
   }
   \label{fig:denni}
\end{figure}
To hydrodynamically model the Type IIP SN~2017gmr, we use the time-implicit
   Lagrangian radiation hydrodynamics code {\sc Crab} which integrates the set
   of spherically symmetric hydrodynamic equations including self-gravity,
   and a radiation transfer equation in the gray approximation
   \citep{Utr_04, Utr_07}.
The explosion of the pre-SN model is initiated by a supersonic piston applied to
   the stellar envelope at the boundary with the collapsing 1.6\Msun core.
The pre-SN model is the hydrostatic non-evolutionary model of a red
   supergiant (RSG) star.
It should be emphasized that the choice of the non-evolutionary model is 
   motivated by the following arguments.
First, the description of the light curve and photospheric velocities of
   SNe~IIP by the spherically symmetric hydrodynamics cannot be attained
   based on the current evolutionary pre-SN models \citep{UC_08}:
   the modification of the density distribution of the hydrogen envelope
   and the smoothing density and composition gradients at the metals/He
   and He/H composition interfaces is needed.
Second, the RSG explosion in three-dimensional (3D) hydrodynamic simulations
   indeed results in smoothing the density and composition gradients
   at the metals/He and He/H composition interfaces \citep{UWJM_17}.
Third, at the final stage of the pre-SN evolution the density distribution
   in the hydrogen envelope can be modified by the acoustic waves excited
   by the vigorous convection at the Ne-burning stage \citep{SQ_14}.
These arguments unavoidably compels us to choose a non-evolutionary RSG
   model in which the structure modification can be implemented by hand
   to make the spherically symmetric hydrodynamic model appropriate for
   the description of the observational data (Figs.~\ref{fig:denmr} and
   \ref{fig:chcom}).
It is noteworthy that the hydrogen abundance in the inner layers of the mixed
   ejecta $X_c \approx 0.05$ is comparable to the value of about 0.03
   produced by the 3D hydrodynamic model of SN~1999em \citep{UWJM_17}.
However, mixing induced by the SN explosion can be affected by particular
   features of the explosion mechanism, e.g., explosion asymmetry, so
   the higher hydrogen abundance in the central zone of the expanding 
   ejecta cannot be ruled out.

We do not solve the complicated optimization problem to construct the optimal
   pre-SN model, instead we rely on the well-studied effects of model
   parameters on the light curve and the photospheric velocity
   \citep[e.g.,][]{GIN_71, Woo_88, Utr_07}.
The optimal pre-SN model that reproduces the major observational data of
   SN~2017gmr (i.e., the light curve and the evolution of the photospheric
   velocity) is found by means of hydrodynamic simulations for an extended
   set of SN parameters and a several options for the energy source at
   the plateau/tail transition, the distribution of radioactive $^{56}$Ni
   being fixed by the nebular spectra.
   
Along with the pre-SN radius, the ejecta mass, the explosion energy, and
   the total amount of radioactive $^{56}$Ni, the extent of its mixing
   in velocity space affects the light curve at the plateau as well.
For SN~2017gmr the extent of $^{56}$Ni mixing is constrained by the \Ha
   emission on day 312 (Section~\ref{sec:asym}).
We find that the outer boundary of $^{56}$Ni ejecta should lie at about
   3300\kms in the freely expanding envelope. 
The overall density and $^{56}$Ni distribution in the ejecta on day 50  
   for model HM-optm (Table~\ref{tab:hydmod}) is shown in Fig.~\ref{fig:denni}.
Note that in the broad range of expansion velocities 6000\kms $< v < 30000$\kms
   the density distribution follows the power law $\rho \propto v^{-8}$.

\begin{table}
\centering
\caption[]{Basic properties of hydrodynamic models$^a$.}
\label{tab:hydmod}
\begin{tabular}{@{} l @{\:} c @{\:} c @{\:} c @{\:} c @{\:} c l @{}}
\hline
\noalign{\smallskip}
 Model & $M_{ej}$ & $E$ & $M_{\mathrm{Ni}}$ & $v_{\mathrm{Ni}}^{max}$
       & $X_c$ & Note \\
       & (\Msun) & ($10^{51}$\,erg) & (\Msun) & (km\,s$^{-1}$)
       & \multicolumn{2}{c}{ } \\
\noalign{\smallskip}
\hline
\noalign{\smallskip}
HM-refm  & 22.0 & 10.2 & 0.16 & 3300 & 0.05 & reference \\
HM-hmix  & 22.0 & 10.2 & 0.18 & 3300 & 0.20 & H mixing \\
HM-lmas  & 14.0 & 10.0 & 0.20 & 3400 & 0.04 & low $M_{ej}$ \\
HM-exni  & 22.0 & 10.2 & 0.23 & 7900 & 0.05 & outer $^{56}$Ni \\
HM-magn  & 22.0 & 10.2 & 0.01 & 3300 & 0.05 & magnetar \\
HM-optm  & 22.0 & 10.2 & 0.11 & 3300 & 0.05 & FM-mechanism \\
\noalign{\smallskip}
\hline
\multicolumn{7}{l}{$^a$ In all models the pre-SN radius $R_0$ is 525\Rsun.} \\
\end{tabular}
\end{table}
\begin{figure}
   \includegraphics[width=\columnwidth, clip, trim=0 19 27 120]{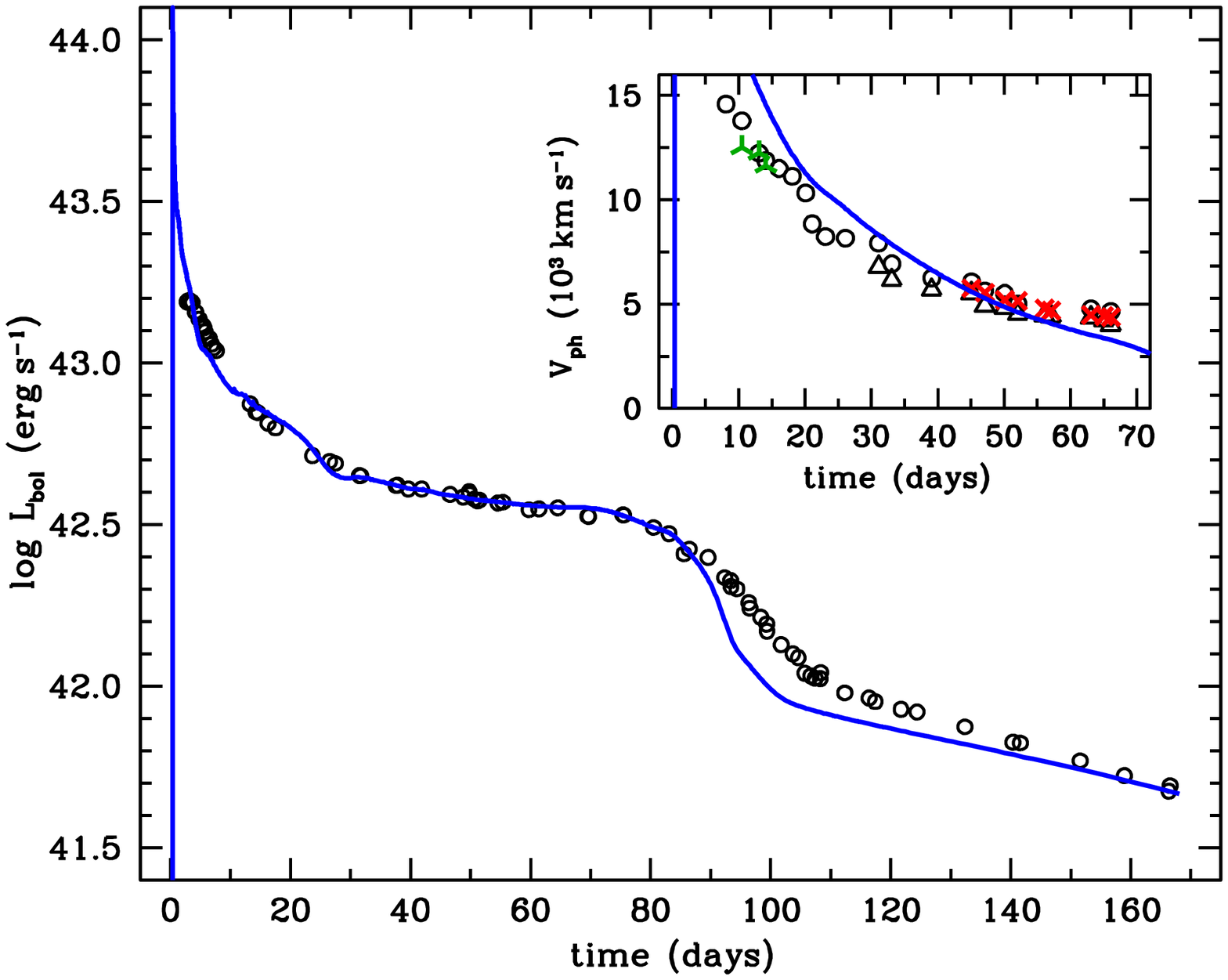}
   \caption{
   The bolometric light curve of model HM-refm (\emph{blue line},
      Table~\ref{tab:hydmod}) overplotted on the observational data
      (\emph{black circles\/}).
   Time is measured in days from core collapse.
   Inset shows the evolution of model velocity at the photosphere level
      defined by the effective optical depth of 2/3 (\emph{blue line\/})
      compared to the photospheric velocities estimated from the \Hb,
      \HeI 5876\A, and \FeII 5169\A lines and the \NaI 5892\A doublet.
   See Section~\ref{sec:bolcrv} and caption of Fig.~\ref{fig:vphcrv}
      for details.
   }
   \label{fig:refm}
\end{figure}
\begin{figure}
   \includegraphics[width=\columnwidth, clip, trim=0 20 47 176]{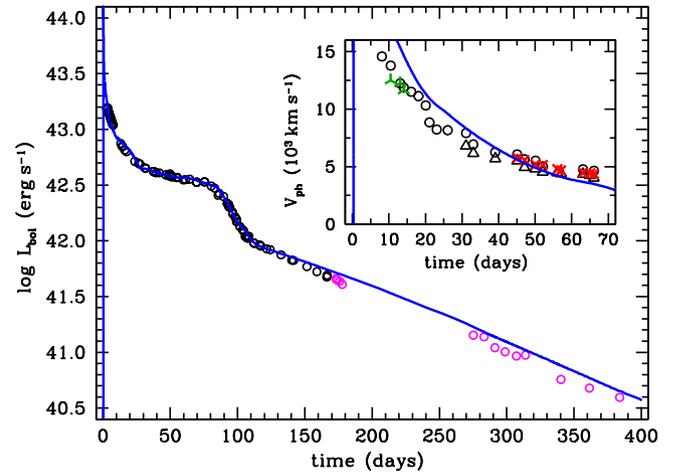}
   \caption{
   The bolometric light curve of model HM-hmix with the enhanced hydrogen
      abundance in the central zone of the ejecta (\emph{blue line},
      Table~\ref{tab:hydmod}) overplotted on the observational data.
   Inset shows the evolution of model velocity at the photosphere level.
   See caption of Fig.~\ref{fig:refm} for details.
   }
   \label{fig:hmix}
\end{figure}
%

\subsection{Plateau/tail transition and luminosity tail}
\label{sec:hydro:optmod} 
The hydrodynamic modelling leads us to the reference model HM-refm
   (Table~\ref{tab:hydmod}) that fits in general the initial luminosity peak,
   the plateau, the late-time radioactive tail, and the photospheric velocities
   of SN~2017gmr (Fig.~\ref{fig:refm}).
However, this model demonstrates the essential deficit in the luminosity at
   the plateau/tail transition in the range of $90-140$\,days.
The impression is that the plateau/tail transition of SN~2017gmr is overlong
   compared to ordinary SNe~IIP.
Note that the recent hydrodynamic modelling of SN~2017gmr by \citet{GB_20}
   faces the same problem of the model drawback at the plateau/tail transition.

Increasing both the total amount of radioactive $^{56}$Ni and the opacity of
   the inner layers of the ejecta might compensate for the luminosity deficit.
The higher opacity could be related to a more intense inward mixing of
   hydrogen-rich matter as a result of a strong explosion asymmetry
   compared to ordinary SNe~IIP.  
The study of this possibility results in model HM-hmix (Table~\ref{tab:hydmod})
   whose $^{56}$Ni mass of 0.182\Msun and hydrogen abundance in the central
   zone $X_c = 0.2$ are larger than 0.158\Msun and $X_c = 0.05$ of the
   reference model, respectively.
The excellent fit to the observations in the range $t \leq 130$\,days lends
   credibility to this model, although the later steep tail decline is not
   fully reproduced (Fig.~\ref{fig:hmix}).

\begin{figure}
   \includegraphics[width=\columnwidth, clip, trim=0 19 54 250]{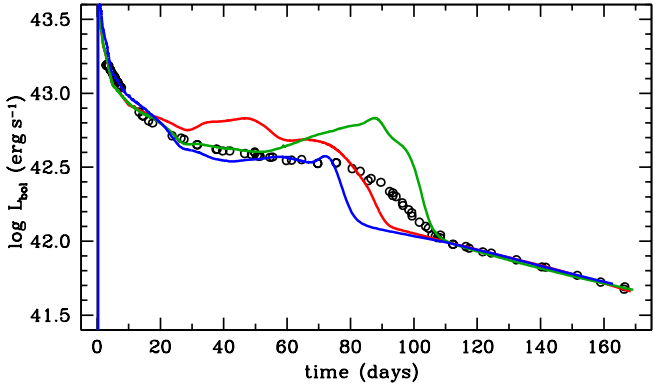}
   \caption{
   The bolometric light curves of three models (Table~\ref{tab:hydmod}) that
      are able to reproduce the observed luminosity tail: model HM-lmas with
      the low ejecta mass (\emph{blue line\/}), model HM-exni with the external
      $^{56}$Ni (\emph{red line\/}), and model HM-magn with the standard
      magnetar luminosity (\emph{green line\/}).
   }
   \label{fig:tail}
\end{figure}

Among other possibilities to resolve the issue of the steep luminosity tail
   one can admit the low ejecta mass that favors a more efficient escape of
   gamma rays from the $^{56}$Co decay.
The appropriate low-mass model HM-lmas (Table~\ref{tab:hydmod}) that fits
   the tail requires the ejecta mass of 14\Msun with the $^{56}$Ni mass
   of 0.2\Msun (Fig.~\ref{fig:tail}, blue line). 
This model expectedly produces too short plateau and should be rejected.

Another possibility is prompted by the model with the external $^{56}$Ni
   ejecta that was successfully applied to SN~2013ej \citep{UC_17}.
The similar hydrodynamic model HM-exni (Table~\ref{tab:hydmod}) for SN~2017gmr
   with the external $^{56}$Ni fits the radioactive tail (Fig.~\ref{fig:tail},
   red line), if all the 0.23\Msun of $^{56}$Ni resides in the velocity range
   of $6000-8000$\kms.
The drawback of this model is a pronounce luminosity excess at the plateau
   in the interval of $30-80$\,days and the luminosity deficit at the
   plateau/tail transition.
The external $^{56}$Ni thus cannot resolve the problems of the anomalous
   light curve of SN~2017gmr.

One can consider somewhat exotic model HM-magn in which the magnetar,
   not radiative $^{56}$Ni, determines the luminosity tail.  
With a standard magnetar luminosity evolution $L = L_0/(1 + t/t_0)^2$
   the tail can be described with the following parameter values:
   $L_0 = 4.47\times10^{43}$\ergs and $t_0 = 19.07$\,days.
The rest of model parameters are in Table~\ref{tab:hydmod}.
The apparent drawback of the magnetar model is a large luminosity excess
   at the end of the plateau (Fig.~\ref{fig:tail}, green line).
The magnetar model in a simple version thus cannot resolve simultaneously
   the issue of the plateau/tail transition and  of the steep luminosity tail.

We briefly address a possible role of the CS interaction.
Although this mechanism potentially is able to provide the steep luminosity
   tail; however, in this case it is highly unlikely.
The point is that the interaction power of $\sim$10$^{42}$\ergs required
   at about 100\,days is released in the outer layers of the ejecta that
   would be inevitably accompanied by a strong broad \Ha emission with
   the luminosity of $\sim$10$^{41}$\ergs and a specific line profile
   lacking the absorption component.
This is not the case for SN~2017gmr \citep[cf.][]{ASV_19}, so the significant
   contribution of the spherical CS interaction is ruled out.
To overcome this problem, one might admit that the CS matter is arranged in
   the form of a dense equatorial ring that is overtaken by the SN envelope
   \citep[e.g.,][]{CD_94, SMC_15}.
In that case one expects that the forward and reverse shocks are submerged
   inside the envelope, so the released radiation creates a quasi-spherical
   photosphere.
This scenario, however, should reveal strong effects of the non-spherical
   interaction in the early \Ha profile.
Instead, the spectra show the usual steady evolution --- a characteristic of
   spherical SNe~IIP.
The CS interaction as a major energy source at the early luminosity tail thus
   should be rejected.

\begin{figure}
   \includegraphics[width=\columnwidth, clip, trim=0 20 47 176]{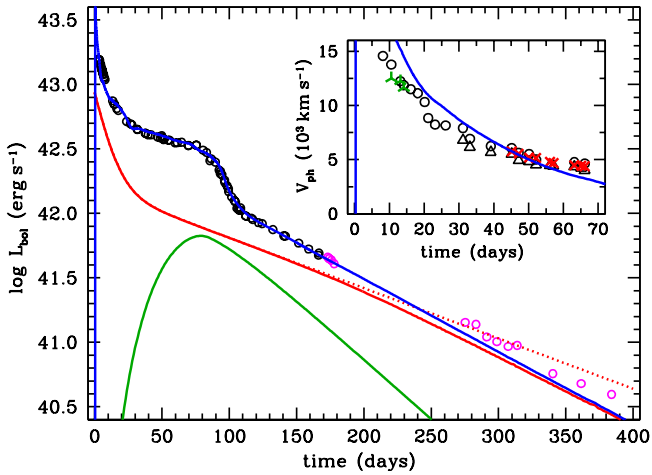}
   \caption{
   The bolometric light curve of model HM-optm (\emph{blue line},
      Table \ref{tab:hydmod}) overplotted on the observational data
      (\emph{black and magenta circles\/}).
   A good fit to the observations is obtained with the additional energy
      source applied to the internal boundary of the ejecta
      (\emph{green line\/}) and the energy deposition of gamma rays
      by radioactive decay of $^{56}$Ni (\emph{solid red line\/}).
   \emph{Dotted red line\/} is the total power of the radioactive decay.
   Inset shows the evolution of the photospheric velocity in the same way
      as in Fig.~\ref{fig:refm}.
   }
   \label{fig:optm}
\end{figure}
%

\subsection{Central energy source?}
\label{sec:hydro:prop}
An interesting possibility to resolve the issue of the post-plateau light curve
   involves an additional energy source related to the fallback on the magnetar.
The conjecture looks quite sensible by two reasons: first, the fallback
   accretion flow always accompanies the core-collapse supernovae
   \citep{Col_71, Che_89}, and second, the magnetar formation could be
   a natural outcome of the magneto-rotational explosion mechanism that
   is required to provide the enormous explosion energy of
   $\sim$10$^{52}$\,erg.
The extra energy source involving both the fallback and the magnetar we dub
   ``FM-mechanism'', for short.
The dense environment associated with the fallback inhibits the magnetic
   dipole radiation of the magnetar, whereas the FM-mechanism could release
   rotational energy in the propeller regime that operates when the Alfven
   radius $R_m$ exceeds the corotation radius $R_c$ and both of these radii
   are less than the radius of the light cylinder \citep{IS_75, Sha_75}.
Conditions required for the FM-mechanism to operate successfully can be 
   illustrated by the case of the unusual supernova ASASSN-15nx with the
   luminosity decreasing in the range of $5\times10^{42}$\ergs to
   $5\times10^{41}$\ergs between days 100 and 200 \citep{Chu_19}.
In that case the light curve has been modelled in terms of the FM-mechanism
   assuming a spherical fallback with the accretion rate
   $\dot{M} \sim 10^{-3}\,M_{\sun}$\,yr$^{-1}$ onto the magnetosphere of
   the neutron star with the magnetic momentum
   $\mu \sim 5\times 10^{31}$\,G\,cm$^3$, and the rotation period
   $p \sim 10^{-2}$\,s.
These values could be applicable for the FM-mechanism in the case of SN~2017gmr,
   although some complications could arise due to the unknown specific angular
   momentum of the fallback material.

We will constrain the rate of the power release by the FM-mechanism, $L_c(t)$,
   using the hydrodynamic modelling.
To this end we impose the power $L_c(t)$ at the inner boundary of the
   computational domain immediately after the explosion.
The model HM-optm (Table \ref{tab:hydmod}) with the $^{56}$Ni mass of 0.11\Msun
   reproduces both the observed plateau/tail transition and the luminosity
   tail decline (Fig. \ref{fig:optm}) for the adopted luminosity $L_c(t)$
   shown in the same plot.
The required evolution of the additional energy source is characterized by
   a slow rise toward the maximum at about 80\,days and an exponential decline
   later on.
In fact, the exponential behavior is constrained only to the stage of
   $t < 180$\,days; we retain the later exponential behavior simply to minimize
   a number of parameters, although the preferred option is zero contribution
   of the extra energy source to the luminosity on day 312.

At first glance the adopted luminosity evolution of the FM-mechanism looks
   highly artificial.
In fact, it is not, because the fallback accretion rate at the early stage is
   high enough to shrink the magnetosphere so strongly that the inequality
   $R_m < R_c$ is fulfilled thus turning off the propeller regime.
In this case the gravitational energy of the fallback accretion flow is
   released in the close vicinity of neutron star producing the neutrino
   luminosity \citep{Che_89}.

It should be emphasized that the conjecture about the extra energy source
   is viable so long as the conclusion on the fast decline of the observed
   luminosity tail remains valid.

\begin{figure}
   \includegraphics[width=\columnwidth, clip, trim=0 19 54 245]{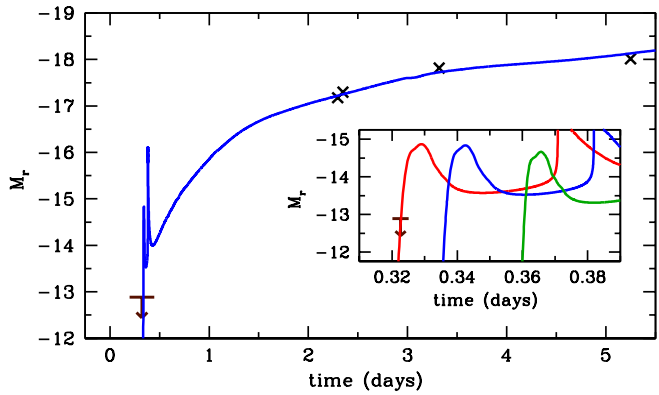}
   \caption{
   Rising part of the $r$-band light curve during the first 5.5\,days for
      model HM-optm (\emph{blue line\/}) compared to observational
      data \citep[][\emph{black crosses\/}]{ASV_19}.
   The \emph{brown arrow} shows the observational upper limit taken by
      \citet{ASV_19} about two days prior to SN discovery. 
   Inset shows sensitivity of the rising part of the $r$-band light curve
      for the models with different ejecta masses at $E/M_{ej} =$ const:
      model with $M_{ej} = 21$\Msun (\emph{red line\/}), model HM-optm
      with $M_{ej} = 22$\Msun (\emph{blue line\/}), and
      model with $M_{ej} = 24$\Msun (\emph{green line\/}).
   Model with $M_{ej} = 21$\Msun imposes the lower limit of the ejecta mass
      (Section~\ref{sec:hydro:mass}).
   }
   \label{fig:rise}
\end{figure}
\begin{figure}
   \includegraphics[width=\columnwidth, clip, trim=0 28 4 27]{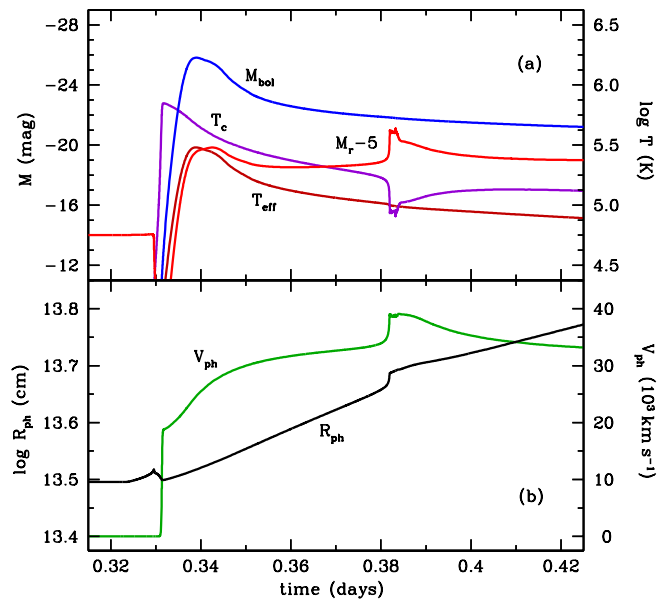}
   \caption{
   The physics of the double peak structure in the $r$-band maximum occurring
      during the first two hours after the shock breakout.
   Panel (a): the first maximum in the $r$-band light curve (\emph{red line\/})
      at $t \sim 0.34$\,days is related to the maximum of the bolometric
      luminosity (\emph{blue line\/}), while the second maximum at
      $t \sim 0.38$\,days is related to the formation of the thin shell
      accompanied by the drop of the color temperature (\emph{magenta line\/}).
   Note that both the bolometric luminosity and the effective temperature
      (\emph{Indian red line\/}) monotonically decrease around the second
      maximum.
   Panel (b): the behavior of the photospheric radius (\emph{black line\/})
      and velocity (\emph{green line\/}) in the vicinity of the second maximum
      reflects the thin shell formation.
   }
   \label{fig:dpeak}
\end{figure}
%

\subsection{Constraining the ejecta mass}
\label{sec:hydro:mass}
The adequate hydrodynamic model is constructed  by fitting the photometric
   and spectroscopic observations of the object under study.
The important physical parameters --- the initial radius $R_0$, the ejecta
   mass $M_{ej}$, and the explosion energy $E$ --- can be reliably estimated
   from the detailed observations of the whole outburst, particularly,
   at both the shock breakout and the plateau/tail transition.
Photometric data of SN~2017gmr are well defined at both epochs and are
   sufficiently comprehensive to construct the bolometric light curve
   \citep{ASV_19}.
The rising part of the $r$-band light curve after the shock breakout is
   strengthened by the upper limit in $r$-band obtained two days before
   the SN discovery \citep{ASV_19} in constraining the basic parameters
   of hydrodynamic model.

Model HM-optm excellently fits the early $M_r$ magnitudes and
   satisfies the upper limit provided the explosion occurs at
   MJD 57997.89 (Fig.~\ref{fig:rise}).
It is noteworthy that during the first several days the SN spectrum is
   essentially the Planck function that permits us to adequately calculate
   the $r$-band magnitude with the radiation hydrodynamics code {\sc Crab}.   
Interestingly, the flux in $r$-band shows a double peak structure that is
   to our knowledge has never been noticed either observationally or
   numerically.
In this particular model the double peak phenomenon is related to the
   formation of the opaque thin dense shell at about 0.38\,days after
   the collapse and $\sim$1 hour after the shock breakout due to the
   sweeping of the external layers into the thin shell by the pressure
   of the escaping radiation flux. 
To illustrate the point we show the evolution of related values, viz.,
   the absolute bolometric magnitude $M_{bol}$, the absolute $r$-band
   magnitude $M_r$, the effective temperature $T_{eff}$, the color
   temperature $T_c$, the photospheric velocity $v_{ph}$, and the photospheric
   radius $R_{ph}$ (Fig.~\ref{fig:dpeak}).
It is noteworthy that the $r$-band flux shows a jump at $\sim$1 hour after
   the shock breakout, whereas the bolometric flux does not. 
This suggests that the variation of the $r$-band flux is essentially related
   to that in the color temperature (Fig.~\ref{fig:dpeak}a).
The formation of the thin opaque shell is accompanied by its radiative
   acceleration clearly seen in the behavior of the velocity at the photosphere
   (Fig.~\ref{fig:dpeak}b).
The double peak structure in the $r$ (in other bands as well) light curve is
   an interesting phenomenon that could be used in future as an additional tool
   in order to constrain SN~IIP parameters.

\begin{figure}
   \includegraphics[width=\columnwidth, clip, trim=2 18 27 28]{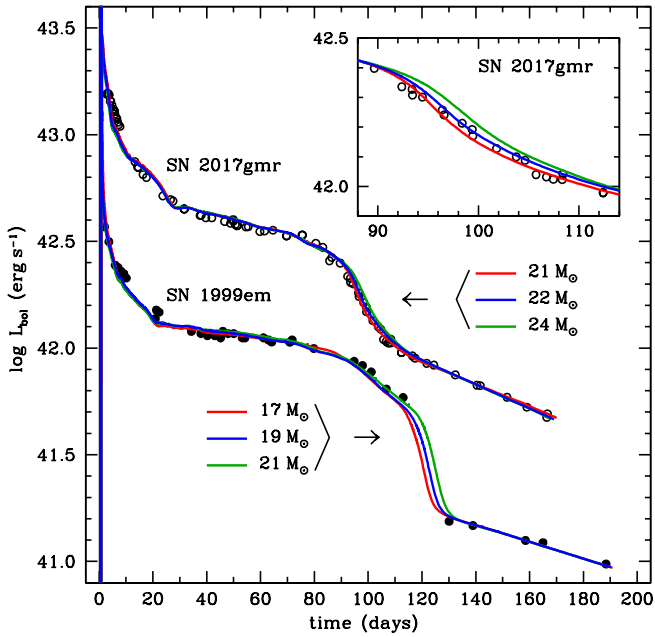}
   \caption{
   Sensitivity of the bolometric light curve to the ejecta mass variation
      at $E/M_{ej} =$ const for SN~1999em and SN~2017gmr.
   The bolometric light curve of SN~1999em was taken from \citet{ECP_03}.
   In the case of SN~2017gmr the sensitivity to the mass variation is
      relatively weak, because the plateau/tail transition is affected by
      the additional luminosity of the central source.
   Inset shows that even in this case the models with different ejecta masses
      are clearly distinguished.
   }
   \label{fig:emass}
\end{figure}

The plateau length $t_p$ of hydrodynamic models of SNe~IIP is known to be
   almost independent of the ejecta mass given the constant ratio $E/M_{ej}$.
For example, in the case of the normal IIP SN~1999em it shows only weak
   dependence on the explosion energy $t_p \propto E^{\,-0.18}$ and the
   pre-SN radius $t_p \propto R_{0}^{\,0.10}$ \citep{Utr_07}.
Such a behavior of hydrodynamic models produces the impression of a degeneracy
   with respect to the model parameters.
However, in the rigorous sense the degeneracy on the ejecta mass is absent.
We demonstrate this fact for SN~1999em with auxiliary hydrodynamic models
   in which $E/M_{ej} =$ const and the calculated bolometric luminosity
   is secured at the observed plateau luminosity by the appropriate choice of
   the pre-SN radius.
These models with the different ejecta masses deviated from the value
   of 19\Msun by $\pm2$\Msun show significant difference at the plateau/tail
   transition (Fig.~\ref{fig:emass}).
This numerical experiment suggests that the distributions of the major elements
   and $^{56}$Ni are fixed in the ejecta (Figs.~\ref{fig:chcom} and
   \ref{fig:denni}).
Any variation of these distributions results in the deformation of the model
   light curve at the end of the plateau and the plateau/tail transition,
   which is inconsistent with observations.
It is noteworthy that the uncertainty of the ejecta mass becomes significant
   ($>$10\%), if the date of the explosion is fixed with an error worse than
   $\sim$2\,days. 

In the special case of SN~2017gmr with the additional central energy source
   the dependence on the ejecta mass is weaker, yet even in this case we are
   able to distinguish between the models with the ejecta masses of 21\Msun,
   22\Msun, and 24\Msun (Fig.~\ref{fig:emass}, inset).
It is remarkable that the upper limit in $r$-band taken two days before
   the SN discovery imposes serious constraint on the ejecta mass.
Its physical meaning can be explained by the following reasonings.
Using the approximate formulae relating the physical parameters to the
   observable properties of hydrodynamic models \citep{Utr_07} with
   $E/M_{ej} =$ const and the fixed bolometric luminosity at the plateau,
   we find that $R_{0} \propto M_{ej}^{\,-1.51}$.
In other words, the lower the ejecta mass is, the larger the pre-SN radius is,
   and, consequently, the larger the characteristic expansion time is.
The latter results in the longer time interval between the shock breakout
   and the epoch of the SN discovery at $M_r = - 17.18$\,mag.
This implies that there is a lower limit of the ejecta mass determined by
   the interval between the time of the upper limit in $r$-band and the
   discovery epoch.
This interval is exactly realized in the hydrodynamic model with the ejecta
   mass of 21\Msun (Fig.~\ref{fig:rise}, inset).
Model HM-optm with the ejecta mass of 22\Msun excellently fits
   both the upper limit in $r$-band and the early $M_r$ magnitudes
   (Fig.~\ref{fig:rise}), and the observed bolometric light curve as
   a whole (Fig.~\ref{fig:optm}).

In addition to the lower limit of the ejecta mass of 21\Msun, we can estimate
   the uncertainty in the derived SN parameters by varying the model parameters
   around the optimal model.
The uncertainty of 1.4\,Mpc in the SN distance and the uncertainty of about
   0.06\,mag in the reddening $E(B-V)$ \citep{ASV_19} imply nearly 20 per cent
   uncertainty in the bolometric luminosity.
The scatter in the plot of the photospheric velocity versus time
   (Fig.~\ref{fig:optm}, inset) suggests the uncertainty of about 7 per cent
   in the photospheric velocity.
We estimate the maximal uncertainty of the plateau length as 4\,days, i.e.
   4 per cent of the plateau duration.
With these uncertainties of observables, we have the initial radius of
   $525\pm176$\Rsun, the ejecta mass of $22_{-1}^{+2}$\Msun, the explosion
   energy of $(10.2\pm0.83)\times10^{51}$\,erg, and the total $^{56}$Ni mass
   of $0.110\pm0.023$\Msun.

\section{Asymmetry of $^{56}$Ni ejecta}
\label{sec:asym}
The triple-peaked \Ha profile in the spectrum on day 312 \citep{ASV_19}
   suggests a non-spherical line-emitting region.
Following a concept employed for the Type IIP SN~2004dj \citep{CFS_05},
   the Type IIP SN~2016X \citep{UC_19}, and the Type II SN~2010jp
   \citep{SCB_12}, we attribute the \Ha asymmetry to the bipolar $^{56}$Ni
   ejecta embedded into a spherical envelope. 
For SN~2017gmr the $^{56}$Ni distribution is represented by three collinear
   homogeneous spheres: central, front, and rear.
The radii, the shifts, and the masses of components are found via the fit of
   the \Ha line profile for the optimal inclination angle. 
The latter is constrained by relying on the linear polarization
   $p = 0.37\pm0.04$\% on day 136 \citep{NCP_19}.
The density distribution in the envelope is set analytically
   as $\rho(v) = \rho_0(v_0/v)^{0.5}/[1 + (v/v_0)^{7.5}]$ with
   $\rho_0$ and $v_0$ specified by the ejecta mass of 22\Msun and
   the explosion energy of $10^{52}$\,erg.
The adopted density distribution in the outer layers $\rho \propto v^{-8}$ is
   well consistent with the hydrodynamic model HM-optm
   (Fig.~\ref{fig:denni}). 

The energy deposition by gamma-rays is treated in a single flight
   approximation with the absorption coefficient
   $k_{\gamma} = 0.03(1 + X)$\,cm$^2$\,g$^{-1}$ \citep{KF_92}, where
   $X$ is the hydrogen mass fraction assumed to be uniform in the ejecta.
We adopt $X = 0.5$ to allow for the synthesized helium.
Positrons from $e^-$-capture channel deposit their kinetic energy on-the-spot.
The ionization rate by the Compton and secondary electrons is calculated
   with the energy fractions spent on ionization, excitation, and heating
   according to \citet{XMOR_92}.
The recombination rate corresponds to the total recombinations on levels
   $n > 2$ assuming the electron temperature of 5000\,K.
The adopted recombination regime approximately allows for the ionization
   from the second level by the recombination Balmer continuum and by the
   two-photon hydrogen continuum.
The recombination \Ha emissivity corresponds to case C (opaque Balmer lines)
   which implies that each recombination onto levels $n > 2$ ends up with
   the \Ha quanta emission. 

\begin{figure}
   \includegraphics[width=\columnwidth, clip, trim=10 19 12 264]{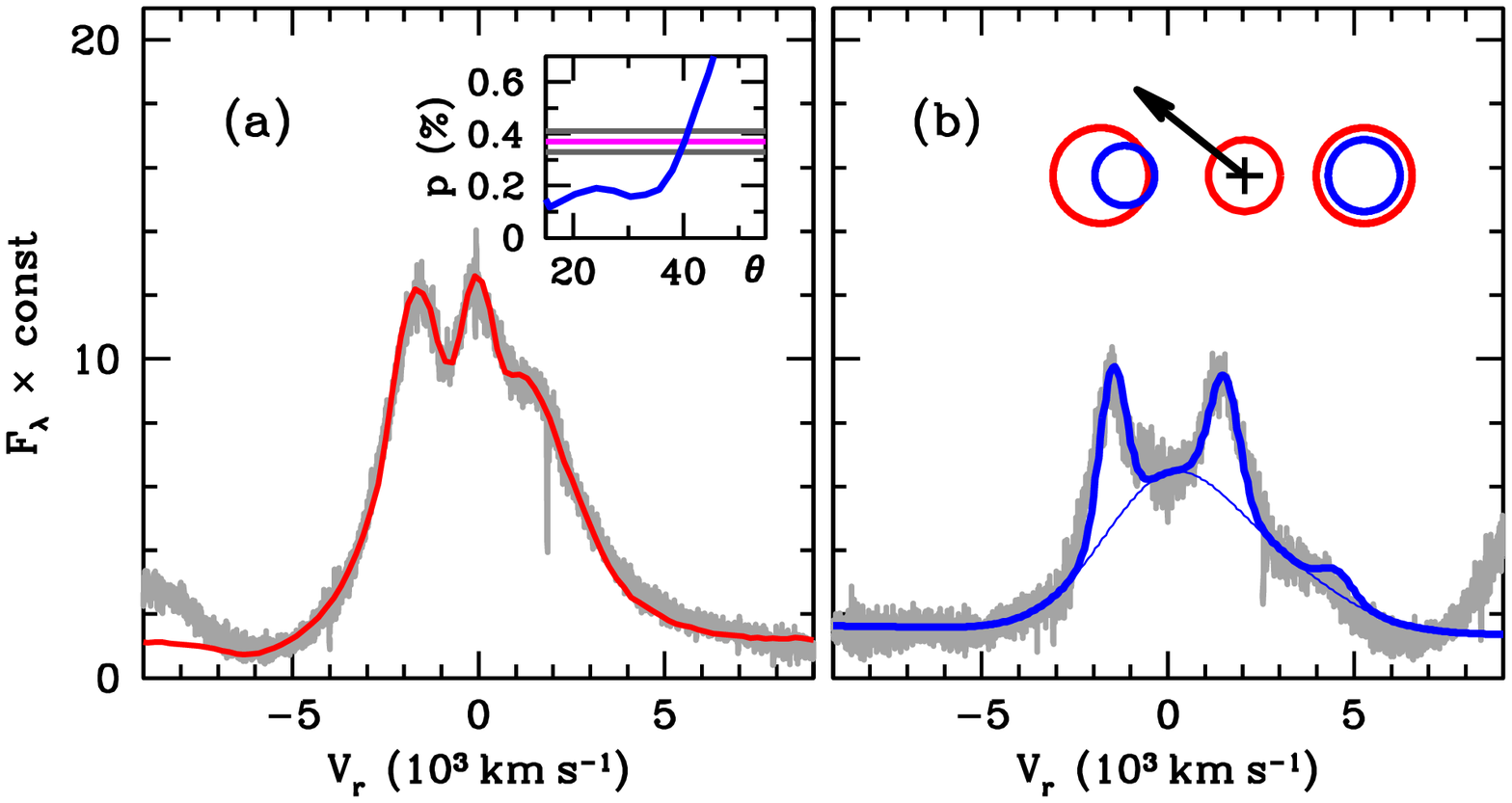}
   \caption{
   Asymmetry in the \Ha and [\OI] 6300, 6364\A line profiles
      taken on day 312 \citep[][\emph{gray lines\/}]{ASV_19}. 
   Panel (a): the model \Ha line is given by \emph{red line}.
   Inset shows the computed polarization as a function of inclination angle
      (\emph{blue line\/}) and the observed polarization degree
      (\emph{horizontal magenta line\/}) along with the $\pm\sigma$ lines
      (\emph{horizontal gray lines\/}) \citep{NCP_19}.
   Panel (b): \emph{thick blue line} represents the resultant model
      [\OI] 6300, 6364\A line and \emph{thin blue line} corresponds
      to the extended spherical emissivity component.
   Cartoon shows the spherical and bipolar components of the radioactive
      $^{56}$Ni distribution (\emph{red circles\/}) and the bipolar
      components of the [\OI] emissivity (\emph{blue circles\/}).
   Arrow indicates the direction towards observer.
   }
   \label{fig:asym}
\end{figure}

The inclination angle is found using the iterative procedure starting with
   a certain $^{56}$Ni configuration: $^{56}$Ni $\rightarrow$ \Ha $\rightarrow$
   polarization $\rightarrow$ $^{56}$Ni etc.
The polarization computation \citep{Chu_06} is based on the Monte Carlo
   technique that follows the history of photons created by a central
   spherical source with a subsequent Thomson multiple scattering in
   the non-spherical distribution of electrons produced by the
   bipolar $^{56}$Ni ejecta.
The found inclination angle is $\theta = 40^{\circ}$ (Fig.~\ref{fig:asym}a)
   with the radii $v_r$, the shifts $v_s$, and the masses of $^{56}$Ni
   components given in Table~\ref{tab:ni}.
The bipolar components are not completely symmetrical: rear component has
   the lower bulk velocity and the lower mass compared to the front component.
For the $^{56}$Ni mass of 0.11\Msun the model \Ha luminosity of
   $3\times10^{39}$\ergs on day 312 coincides with the observed value of
   $3.2\times10^{39}$\ergs estimated from the flux calibrated spectrum.
This choice of the $^{56}$Ni mass is consistent also with the light curve 
   of model HM-optm (Fig.~\ref{fig:optm}).

\begin{table}
\centering 
\caption[]{Parameters of $^{56}$Ni components.}
\label{tab:ni}
\begin{tabular}{l c c c}
\hline
\noalign{\smallskip}
Component & $v_s$ & $v_r$  & $M_{\mathrm{Ni}}$ \\
          & \multicolumn{2}{c}{(km\,s$^{-1}$)} & (\Msun) \\
\noalign{\smallskip}
\hline
\noalign{\smallskip}
 front   & 2400  &  900  &  0.059 \\
 rear    & 2100  &  900  &  0.025 \\
 central &  0    &  700  &  0.026 \\
\noalign{\smallskip}
\hline
\end{tabular}
\end{table}
A reliable modelling of the thermal state of the [\OI] doublet-emitting
   region is beyond reach given the significant role of the cooling by CO and
   SiO in the oxygen-rich matter \citep{LD_95}.
To get idea of the asymmetry of the oxygen line-emitting region, we decompose
   [\OI] 6300, 6364\A doublet using spherical Gaussian components:
   central, front, and rear assuming the inclination angle of 40$^{\circ}$.
The components are specified by the normalized emissivity
   $j = A\exp{[-(u/b)^2]}$, where $u$ is a velocity distance from the center
   of the component with a certain velocity shift $v_s$
   (Table \ref{tab:oxy}, Fig.~\ref{fig:asym}b).
The bipolar components of the oxygen emissivity in [\OI] 6300, 6364\A
   doublet are rather similar to those of $^{56}$Ni (Fig.~\ref{fig:asym}b,
   inset).
This could be interpreted in two ways: either oxygen components reflect
   the distribution of the electron temperature due to the bipolar $^{56}$Ni
   in a spherical oxygen-rich gas, or the oxygen ejecta have essentially
   bipolar structure. 
The absence in the oxygen 6300\A line narrow counterpart related to the
   central $^{56}$Ni component favors the bipolar oxygen distribution.

A conjecture that the [\OI] doublet originates from the toroidal
   oxygen distribution might be conceivable.
However, the modelling shows that for any parameters the toroidal structure
   fails to reproduce the observed [\OI] doublet profile.
The toroidal oxygen distribution is thus ruled out.

\begin{figure}
   \includegraphics[width=\columnwidth, clip, trim=7 20 29 308]{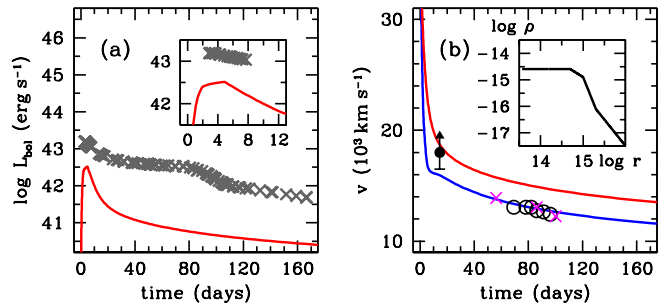}
   \caption{
   Observational effects of the CS interaction for the ejecta mass of 22\Msun
      and the explosion energy of $6\times10^{51}$\,erg.
   Left panel shows the computed bolometric luminosity due to only the
      CS interaction (\emph{red line\/}) compared to the observed bolometric
      luminosity (\emph{gray crosses\/}).
   Inset shows a zoom in of the first ten days that demonstrates a small
      contribution of the CS interaction to the SN luminosity at the stage
      $t < 2$\,days. 
   Right panel shows the evolution of the CDS velocity (\emph{blue line\/})
      and the maximal velocity of the unshocked ejecta (\emph{red line\/}).
   Filled circle corresponds to the lower limit of the maximal ejecta velocity
      inferred from the \HeI 10830\A absorption on day 14.48.
   Crosses correspond to the NHVA inferred from the \Ha profile, while
      open circles are the NHVA inferred from the \HeI 10830\A line.
   Inset shows the corresponding CS density distribution.
   }
   \label{fig:csm}
\end{figure}
\begin{table}
\centering 
\caption[]{Parameters of [\OI] doublet components.}
\label{tab:oxy}
\begin{tabular}{l c c c}
\hline
\noalign{\smallskip}
Component & $v_s$ & $b$ & $A$ \\
          & \multicolumn{2}{c}{(km\,s$^{-1}$)} & \\
\noalign{\smallskip}
\hline
\noalign{\smallskip}
 front   & 2000  &  500  &  1.0 \\
 rear    & 2000  &  600  &  0.3 \\
 central &  0    &  600  &  0.047 \\
\noalign{\smallskip}
\hline
\end{tabular}
\end{table}
%

\section{Circumstellar interaction}
\label{sec:csi}
The narrow \Ha emission revealed by the early spectra of SN~2017gmr
   \citep{ASV_19} on days 1.5 and 2.3 and disappeared 3\,days later suggests
   the presence of a dense confined CS shell similar to that of SN~2013fs
   \citep{YPG_17, BSW_18}.
The hydrodynamic interaction of the ejecta with the CS shell should result in
   the additional optical luminosity powered by the forward and reverse shocks.
Another consequence of the CS interaction is the formation of a thin cold dense
   shell (CDS) between two shocks.
The latter is observed in some SNe~IIP as a narrow high-velocity
   ($\sim$10$^4$\kms) absorption (NHVA) in the \Ha, \HeI 10830\A, and possibly
   \Hb lines at about $50-100$\,days \citep{CCU_07}.
The optical and near infrared spectra of SN~2017gmr indeed reveal distinctive
   NHVA in the \Ha and \HeI 10830\A lines with velocities decreasing in
   the range $14000-12000$\kms between days 50 and 100.
We identify this feature with the CDS partially fragmented due to
   Rayleigh-Taylor instability \citep{CCU_07}.
The manifestation of the CS interaction as the luminosity excess and the NHVA
   can be used to constrain the parameters of the SN ejecta and CS shell.

We treat the CS interaction based on a thin shell approximation \citep{Che_82}.
The interaction model was described earlier \citep{Chu_01} and here we recap
   only the essential points.
The CDS dynamics is computed using Runge-Kutta 4-th order solver
   \citep{PTVF_07}. 
The shock radiative cooling time $t_c$ at a certain moment is determined
   assuming the electron-ion equilibration with the postshock density four
   times greater that the preshock density.
This description is not valid at the very early stage, $t < 10$ hours after
   the shock breakout, when the radiative precursor strongly accelerates
   the preshock gas thus diminishing the viscous shock.
The forward and reverse shock luminosity is approximated as $L_k/(1 + t_c/t)$,
   where $L_k$ is the shock kinetic luminosity.
The interaction optical luminosity is equal to the X-ray luminosity absorbed
   by the unshocked ejecta and the CDS.
The density of the homologously expanding ejecta is set as before
   $\rho = \rho_0(v_0/v)^{0.5}/[1 + (v/v_0)^{7.5}]$.
The density distribution of the confined CS shell is adopted to be uniform
   in the range of $r <  5\times10^{14}$\,cm with a drop $\rho \propto 1/r^4$
   in the range of $(1-2)\times10^{15}$\,cm and the steady wind
   $\rho \propto 1/r^2$ in the outer zone $r >  2\times10^{15}$\,cm
   (Fig.~\ref{fig:csm}b, inset).
The adopted CS shell extent is in line with that for SN~2013fs,
   $(0.4-1)\times10^{15}$\,cm \citep{YPG_17}.

Our strategy is to find a model that minimizes the CS luminosity and
   meets the kinematic requirements imposed by the NHVA.
The luminosity and the kinematic properties of the CDS depend on the CS density
   and the density $\rho(v)$ of the external layers of the SN ejecta.
For a given ejecta mass of 22\Msun the latter is determined by the explosion
   energy that can be found from the luminosity and the kinematic constraints.
We find that the preferred explosion energy is of $6\times10^{51}$\,erg,
   whereas the mass of the confined CS shell is of $4\times10^{-3}$\Msun.
Remarkably the latter value is comparable to the mass estimate for the confined
   dense CS shell in SN~2013fs \citep{YPG_17}.
This case produces a moderate interaction luminosity with the maximal 
   contribution of about 18\% in the range of $2-7$\,days (Fig.~\ref{fig:csm}a)
   that compensates a small deficit in the bolometric luminosity of the
   hydrodynamic model HM-optm at this stage (Fig.~\ref{fig:optm}).
Simultaneously, the CS interaction model meets the kinematic constraints
   imposed by the early \HeI 10830\A absorption and the NHVA of \Ha and
   \HeI 10830\A (Fig.~\ref{fig:csm}b).
  
The tension between the explosion energy of $6\times10^{51}$\,erg suggested
   by the CS interaction model and that of $10^{52}$\,erg implied by the
   hydrodynamic model should not be considered as an irreducible one.
Given the simplicity of the CS interaction model, this disparity indicates
   that the realistic hydrodynamic model should include the shock wave
   propagation in the CS shell, which however would require a more complex
   hydrodynamic approach in order to treat an essentially 3D physics related
   to the CDS formation and the Rayleigh-Taylor instability.

\begin{figure}
   \includegraphics[width=\columnwidth, clip, trim=4 24 28 27]{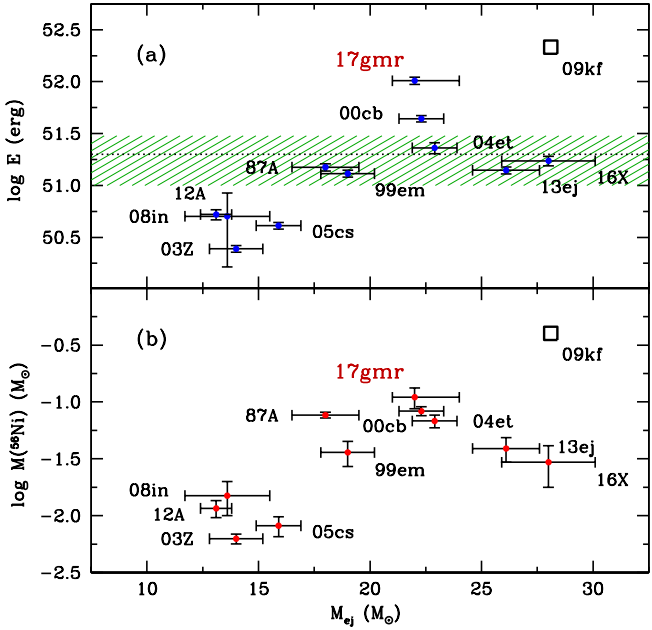}
   \caption{
   Explosion energy (Panel (a)) and $^{56}$Ni mass (Panel (b))
      versus ejecta mass for SN~2017gmr and eleven other
      core-collapse SNe \citep{UC_19}.
   Dotted line in Panel (a) is the upper limit of the explosion energy of
      $2\times10^{51}$\,erg for the neutrino-driven mechanism \citep{Jan_17}
      with the uncertainty of about $\pm 10^{51}$\,erg\protect\footnotemark\
      shown by the shaded \emph{green band}.
   }
   \label{fig:ennims}
\end{figure}
\footnotetext{H.-Th.~Janka, private communication.}  

\section{Discussion and Conclusions}
\label{sec:disc}
%
\begin{table}
\centering
\caption[]{Hydrodynamic models of Type IIP supernovae.}
\label{tab:sumtab}
\begin{tabular}{@{ } l  c  c @{ } c @{ } c @{ } c  c @{ }}
\hline
\noalign{\smallskip}
 SN & $R_0$ & $M_{ej}$ & $E$ & $M_{\mathrm{Ni}}$ 
       & $v_{\mathrm{Ni}}^{max}$ & $v_{\mathrm{H}}^{min}$ \\
       & (\Rsun) & (\Msun) & ($10^{51}$\,erg) & ($10^{-2}$\Msun)
       & \multicolumn{2}{c}{(km\,s$^{-1}$)}\\
\noalign{\smallskip}
\hline
\noalign{\smallskip}
 1987A  &  35  & 18   & 1.5    & 7.65 &  3000 & 600 \\
1999em  & 500  & 19   & 1.3    & 3.6  &  660  & 700 \\
2000cb  &  35  & 22.3 & 4.4    & 8.3  &  8400 & 440 \\
 2003Z  & 230  & 14   & 0.245  & 0.63 &  535  & 360 \\
2004et  & 1500 & 22.9 & 2.3    & 6.8  &  1000 & 300 \\
2005cs  & 600  & 15.9 & 0.41   & 0.82 &  610  & 300 \\
2008in  & 570  & 13.6 & 0.505  & 1.5  &  770  & 490 \\
2009kf  & 2000 & 28.1 & 21.5   & 40.0 &  7700 & 410 \\
2012A   &  715 & 13.1 & 0.525  & 1.16 &  710  & 400 \\
2013ej  & 1500 & 26.1 & 1.4    & 3.9  &  6500 & 800 \\
 2016X  &  436 & 28.0 & 1.73   & 2.95 &  4000 & 760 \\
2017gmr &  525 & 22.0 & 10.2   & 11.0 &  3300 & 640 \\
\noalign{\smallskip}
\hline
\end{tabular}
\end{table}
The paper has been aimed at the study of the non-standard Type IIP SN~2017gmr
   with the emphasis on the unusual light curve at the plateau/tail transition
   and the steep tail decline.
We find that a standard hydrodynamic model of the RSG star exploding with
   the $^{56}$Ni ejecta is not able to describe the plateau/tail transition
   and the very tail for any model parameter set.
In that sense we confirm the conclusion of \citet{GB_20} who demonstrate that
   their hydrodynamical model of SNe~IIP fails to account for the
   plateau/tail transition and the luminosity tail of the SN~2017gmr light
   curve.
The model with the high hydrogen abundance in the central zone of
   the ejecta and the high amount of $^{56}$Ni is able to fit the light curve
   in the range of $t < 130$\,days, although later on the model tail is somewhat
   less steep compared to the observed one.
Yet this kind of hydrodynamic model should be considered as a viable contender
   for the hydrodynamic model with the additional central energy source.

We find also that the light curve of SN~2017gmr can be reproduced in the
   framework of the FM-mechanism which implies a central energy source with
   the specific temporal behavior of the power release.
The extra energy source is attributed in this case to the fallback interaction
   with the magnetar in the propeller regime.
This ad hoc scenario is in line with the extremely high explosion energy of
   $\approx$10$^{52}$\,erg and the bipolar $^{56}$Ni asymmetry, both 
   indicative of the magneto-rotational explosion.
At the moment we are not aware of other SNe~IIP that would show similar behavior
   of the bolometric light curve at the plateau/tail transition and the early
   stage of the luminosity tail.
The another SN~IIP that demonstrates a steep decline of the early tail
   is SN~2013ej \citep{DKV_16}.
Although this behavior has been explained in the model with the external
   $^{56}$Ni \citep{UC_17}, we do not rule out the FM-mechanism as a viable
   alternative.
One should keep eye open on the possibility that the FM-mechanism could
   sometimes manifest itself at the radioactive tail of core-collapse SNe
   including SNe~II varieties and SNe~Ib/c.

The unusual bolometric light curve of SN~2017gmr raises a serious caveat.
The point is that at the plateau/tail transition of SNe~IIP the radiative
   cooling regime changes from the photospheric to the nebular one, which
   is accompanied by the corresponding spectrum transformation.
This poses a question whether the technique for the bolometric flux
   reconstruction that is appropriate for the photospheric stage preserves
   the same accuracy at the nebular stage.
The only case when we have no doubts in this regard is SN~1987A.
In other cases of SNe~IIP some degree of doubt remains. 

The hydrodynamic modelling confirms the earlier suggestion that the high
   luminosity and the fast expansion of the ejecta indicate the high explosion
   energy \citep{ASV_19}.
The explosion energy of $10^{52}$\,erg inferred in our model is twice as large
   compared to $4.6\times10^{51}$\,erg, the value preferred by \citet{GB_20}.
Note, however, that the degeneracy of the light curves for SN~2017gmr admits
   also a model with the larger explosion energy of $\approx$10$^{52}$\,erg
   and, consequently, the ejecta mass larger than 22\Msun \citep{GB_20}.
Despite both hydrodynamic codes {\sc STELLA} \citep{BEB_98, BS_04, BBP_05,
   BRS_06} used by \citet{GB_20} and our code {\sc Crab} \citep{Utr_04, Utr_07}
   produce the similar results for the normal Type IIP SN~1999em
   \citep{BBP_05, Utr_07}, in the case of SN~2017gmr some inconsistency is
   unavoidable because of the different adopted $^{56}$Ni distribution and
   the different pre-SN configuration, namely evolutionary and non-evolutionary
   pre-SN models, respectively.

We have identified the NHVA in the \Ha and \HeI 10830\A lines in the spectra
   between 50\,days and 100\,days and use the velocities of these features
   to constrain the mass of the confined CS shell of
   $\sim$4$\times10^{-3}$\Msun which turns out to be comparable to that of
   SN~2013fs \citep{YPG_17}.
The requirement of the low contribution of the CS interaction luminosity
   combined with the kinematic constraints from the NHVA implies the preferred
   explosion energy of $6\times10^{51}$\,erg that is lower than the energy
   implied by the hydrodynamic model.
The contradiction casts a shadow on the thin shell model and suggests the need
   for the full radiation hydrodynamics treatment of the shock wave
   propagation in the dense CS shell, which however cannot be done by
   the available hydrodynamic code. 

The explosion energy of SN~2017gmr is indeed enormous for SNe~IIP and places
   this event to the category of high-energy SNe~IIP with other two cases of
   SN~2000cb and SN~2009kf (Table~\ref{tab:sumtab}, Fig.~\ref{fig:ennims}).
The explosion energies of these three supernovae exceed the upper limit
   for the neutrino-driven explosion mechanism which, implies that their
   explosions could be related to the rotational energy of the collapsing core.
Unfortunately, spectra of SN~2009kf and SN~2000cb at the late nebular stage
   are lacking, so one cannot say anything about possible asymmetry of
   the $^{56}$Ni ejecta in these supernovae.
It is noteworthy that in both preferred models for SN~2017gmr, HM-hmix and
   HP-optm (Table~\ref{tab:hydmod}), the $^{56}$Ni mass exceeds the amount
   of $^{56}$Ni typical for SNe~IIP (Table~\ref{tab:sumtab},
   Fig.~\ref{fig:ennims}), which is in line with the unusually high explosion
   energy of SN~2017gmr.

\section*{Acknowledgements}
VPU is partially supported by Russian Scientific Foundation grant
   19-12-00229.
This work makes use of observations from the Las Cumbres Observatory global
   telescope network.
The LCO team is supported by NSF grants AST-1911225 and AST-1911151.

\section*{Data Availability}
The data underlying this article will be shared on reasonable request to
   the corresponding author.


\bsp	
\label{lastpage}
\end{document}